\documentclass[]{tPHM2e}
\usepackage{color}

\begin{document}
\doi{10.1080/14786435.20xx.xxxxxx}
\issn{1478-6443}
\issnp{1478-6435}
\jvol{00} \jnum{00} \jyear{2014} 

\markboth{T. Bzdu\v{s}ek and R. Hlubina}{Philosophical Magazine}

\articletype{}

\title{What is the Pairing Glue in the Cuprates? \break Insights from
  Normal and Anomalous Propagators}

\author{T. Bzdu\v{s}ek$^{\rm a,b}$ and R. Hlubina$^{\rm a}$$^\dagger$
  \thanks{$^\dagger$This is an Author's Original Manuscript of an
    article submitted for consideration in the Philosophical Magazine
    \copyright Taylor \& Francis; Philosophical Magazine is available
    online at http://www.tandfonline.com/loi/tphm20.
\vspace{6pt}}
\\
\vspace{6pt}  
$^{\rm a}${\em{Department of Experimental Physics, Comenius University,
Mlynsk\'{a} Dolina F2, 842~48~Bratislava, Slovakia}}; 
$^{\rm b}${\em{Theoretische Physik, ETH-H\"{o}nggerberg, 
CH-8093 Z\"{u}rich, Switzerland}}
\\
\vspace{6pt}
\received{January 2014}}

\maketitle

\begin{abstract}
Both the pairing and the pair-breaking modes lead to similar kinks of
the electron dispersion curves in superconductors, and therefore the
photoemission spectroscopy can not be straightforwardly applied in
search for their pairing glue. If the momentum-dependence of the
normal and anomalous self-energy can be neglected, manipulation with
the data does allow us to extract the gap function $\Delta(\omega)$
and therefore also the pairing glue. However, in the superconducting
cuprates such procedure may not be justified. In this paper we point
out that the pairing glue is more directly visible in the spectral
function of the Nambu-Gor'kov anomalous propagator and we demonstrate
that this function is in principle experimentally observable.
\bigskip


\end{abstract}

\section{Introduction}
\label{sec_introduction}
Although the nature of the normal state of the cuprates remains
enigmatic, their superconducting state is generally believed to be
quite conventional \cite{Tsuei00}. In fact, superconductivity of the
cuprates is a consequence of the global $U(1)$ symmetry breaking
caused by Cooper-pair condensation, i.e. by the same mechanism as in
the conventional low-$T_c$ superconductors. The Cooper pairs are
formed by electrons in a single CuO$_2$ plane and their spin is $S=0$.
A minor modification is that the orbital wave-function of the Cooper
pairs transforms as the so-called $d$-wave under the point group of
the CuO$_2$ plane.  The major open question of the cuprate physics is
what forms the pairing glue which holds the electrons in the Cooper
pairs together, or even whether such a pairing glue is present at all
\cite{Anderson07}.

If we introduce the Nambu spinors $\alpha_{\bf k}^\dagger=(c_{{\bf
    k}\uparrow}^\dagger,c_{-{\bf k}\downarrow})$, all single-particle
properties of the cuprates should be described by the Nambu-Gor'kov
Green's function ${\cal G}_{\bf k}(\tau)=-\langle T\alpha_{\bf
  k}(-i\tau)\alpha_{\bf k}^\dagger\rangle$ \cite[see,
  e.g.,][]{Rickayzen80}.  Note that we work in the imaginary-time
formalism and that ${\cal G}_{\bf k}(\tau)$ is a 2$\times$2 matrix
\[
{\cal G}_{\bf k}(\tau)=\left(\begin{array}{cc}
G_{{\bf k}\uparrow}(\tau) & F_{{\bf k}\uparrow}(\tau)
\\
F^\ast_{{\bf k}\uparrow}(\tau) & -G_{-{\bf k}\downarrow}(-\tau)
\end{array}\right),
\]
which is determined by two functions: by the diagonal Green's function
(normal propagator) $G_{{\bf k}\sigma}(\tau)=-\langle Tc_{{\bf
    k}\sigma}(-i\tau)c_{{\bf k}\sigma}^\dagger\rangle$, as well as by
the off-diagonal Green's function (anomalous propagator) $F_{{\bf
    k}\sigma}(\tau)=-\langle Tc_{{\bf k}\sigma}(-i\tau)c_{-{\bf
    k}-\sigma}\rangle$.  We emphasize that the Green's function ${\cal
  G}_{\bf k}(\tau)$ may be introduced even if the Landau-BCS Fermi
liquid theory does not apply to the cuprates.

The spectral representation of the Green's function is defined as:
\begin{equation}
{\cal G}_{\bf k}(i\omega_l)=\int_{-\infty}^\infty
\frac{dx {\cal A}_{\bf k}(x)}{i\omega_l-x},
\qquad
{\cal A}_{\bf k}(x)=\left(\begin{array}{cc}
A_{{\bf k}\uparrow}(x) & B_{{\bf k}\uparrow}(x)
\\
B^\ast_{{\bf k}\uparrow}(x) & A_{-{\bf k}\downarrow}(-x)
\end{array}\right),
\label{eq:spectral_representation}
\end{equation}
where we have introduced the spectral function of the normal
propagator $A_{{\bf k}\sigma}(x)$, as well as the spectral function of
the anomalous propagator $B_{{\bf k}\sigma}(x)$. In the Lehmann
representation, these functions read as
\begin{eqnarray*}
A_{{\bf k}\sigma}(x)&=&(1+e^{-x/T})\frac{1}{Z}\sum_{m,n}e^{-E_m/T}
\left|\langle m|c^\dagger_{{\bf k}\sigma}|n\rangle\right|^2
\delta(x-E_n+E_m),
\\
B_{{\bf k}\sigma}(x)&=&(1+e^{-x/T})\frac{1}{Z}\sum_{m,n}e^{-E_m/T}
\langle m|c_{{\bf k}\sigma}|n\rangle\langle n|c_{-{\bf k}-\sigma}|m\rangle
\delta(x-E_n+E_m),
\end{eqnarray*}
where $T$ is the temperature. Throughout this paper we set $k_B=1$ and
$\hbar=1$.  

In singlet superconductors with even parity we expect that $A_{{\bf
    k}\uparrow}(x)=A_{{\bf k}\downarrow}(x)=A_{\bf k}(x)$ and $A_{\bf
  k}(x)=A_{-{\bf k}}(x)$.  For the same reason, we expect that
$B_{{\bf k}\uparrow}(x)=-B_{{\bf k}\downarrow}(x)=B_{\bf k}(x)$ and
$B_{\bf k}(x)=B_{-{\bf k}}(x)$. When combined with the property
$B_{{\bf k}\sigma}(x)=B_{-{\bf k}-\sigma}(-x)$ which follows from
hermiticity, we finally find that the anomalous spectral function is
odd, $B_{\bf k}(x)=-B_{\bf k}(-x)$.

Note that the normal spectral function $A_{\bf k}(x)$ is real and
positive, therefore allowing for a probabilistic interpretation. It is
this function which is measured in the angle-resolved photoemission
experiments (ARPES) \cite{Campuzano04}. Moreover, the integral
$N_S(x)=\frac{1}{\cal V}\sum_{\bf k} A_{\bf k}(x)$, where ${\cal V}$
is the normalisation volume, is directly accessible in the tunneling
experiments \cite{Rickayzen80}.

On the other hand, the function $B_{\bf k}(x)$ is in general complex
and therefore it does not have an obvious probabilistic
interpretation. Moreover, $B_{\bf k}(x)$ depends on the phase of the
condensate. At the present time there seem to be no established
experimental procedures which would determine the function $B_{\bf
  k}(x)$.

Thus only one of the two functions $A_{\bf k}(x)$ and $B_{\bf k}(x)$,
which are needed to characterize the single-particle properties of a
superconductor, is currently experimentally accessible.  In
Section~\ref{sec_pair_breaking} we show that, even in conventional
superconductors where the electrons interact via exchange of bosonic
interactions, this lack of information may lead to severe errors in
identifying the pairing glue by ARPES. We demonstrate that both, the
pairing and the pair-breaking interactions, modify the normal spectral
function $A_{\bf k}(x)$ in essentially the same way, making them
indistinguishable in a naive $A_{\bf k}(x)$-based experiment.  A
similar observation has been made very recently in the context of the
cuprates \cite{Park13}: the magnitude of the dispersion kink at 70~meV
in La$_{2-x}$Sr$_x$CuO$_4$ has been found to decrease by only 30\%
between the superconducting sample with $x=0.20$ and $T_c=$32~K and a
non-superconducting overdoped sample with $x=0.30$ and $T_c=$0~K.
From here the authors conclude that these dispersion kinks can not be
caused by the pairing glue of superconductivity.

Within the Eliashberg theory, the two functions $A_{\bf k}(x)$ and
$B_{\bf k}(x)$ are encoded in two functions $Z_{\bf k}(\omega)$ and
$\Delta_{\bf k}(\omega)$. In Section~\ref{sec_dos} we show that, under
certain simplifying assumptions which happen to be satisfied in the
conventional superconductors, the function $\Delta_{\bf k}(\omega)$ which
characterizes the pairing can be quite unexpectedly extracted from the
measurement of only $A_{\bf k}(x)$. In Section~\ref{sec_dos} we also
describe some recent attempts to apply this type of approach to the
cuprate superconductors.

Finally, in Section~\ref{sec_anomalous} we go beyond the $A_{\bf
  k}(x)$-based spectroscopic techniques and we suggest how, at least
in principle, the anomalous spectral function $B_{\bf k}(x)$ can be
determined directly from the current-current correlation function of
the Josephson junctions.

\section{Eliashberg theory: pairing vs. pair-breaking modes}
\label{sec_pair_breaking}

The Eliashberg theory \cite{Rickayzen80} results from a combination of
the Dyson equation $ {\cal G}_{\bf
  k}(i\omega_n)=\left[i\omega\tau_0-\varepsilon_k\tau_3
  -{\hat\Sigma}_{\bf k}(i\omega_n)\right]^{-1} $ with a
self-consistent Born approximation for the $2\times 2$ self-energy
matrix ${\hat\Sigma}_{\bf k}(i\omega)$.  In the electron-phonon case,
this is virtually exact due to the Migdal ``theorem''. For the sake of
simplicity, let us assume that the system under study is isotropic and
particle-hole symmetric. Then we can make the following ansatz for the
self-energy of an $s$-wave pairing state: ${\hat\Sigma}_{\bf
  k}(i\omega_n)=(1-Z_n)i\omega_n\tau_0+\Delta_n Z_n\tau_1$, where the
wave-function renormalisation $Z_n=Z(i\omega_n)$ and the gap function
$\Delta_n=\Delta(i\omega_n)$ are real functions (only) of frequency,
$\tau_0$ is a $2\times 2$ unit matrix, and $\tau_i$ with $i=1,2,3$ are
the Pauli matrices. The diagonal and off-diagonal components of the
resulting Green's function in this simple case read as
\begin{equation}
G_{\bf k}(i\omega_n)=-\frac{i\omega_n Z_n+\varepsilon_k}
{Z_n^2(\omega_n^2+\Delta_n^2)+\varepsilon_k^2},
\qquad
F_{\bf k}(i\omega_n)=-\frac{Z_n\Delta_n}
{Z_n^2(\omega_n^2+\Delta_n^2)+\varepsilon_k^2}.
\label{eq:green_eliashberg}
\end{equation}

In order to illustrate the difference between the pairing and
pair-breaking modes, let us consider the following simple model of
interacting electrons and bosons:
\begin{equation}
H_{\rm int}=\frac{1}{\sqrt {\cal V}}\sum_{{\bf q}\neq{\bf 0}} 
\left[{\rm g}_{0{\bf q}}\rho_{\bf q} A^0_{-{\bf q}}
+{\rm g}_{\perp{\bf q}}{\bf j}_{\bf q}\cdot {\bf A}_{-{\bf q}}\right].
\label{eq:model}
\end{equation}
The first term describes the coupling of strength ${\rm g}_{0{\bf q}}$
between the electron charge density $\rho_{\bf q}=\sum_{{\bf k}\sigma}
c^\dagger_{{\bf k}-{\bf q}/2\sigma}c_{{\bf k}+{\bf q}/2\sigma}$ and a
scalar mode with the propagator $D_0({\bf q},\tau)=\left\langle
TA^0_{-{\bf q}}(-i\tau)A_{\bf q}^0\right\rangle$ and the Fourier
transform $D_0({\bf q},\omega_n)$. The second term describes the
coupling of strength ${\rm g}_{\perp{\bf q}}$ between the electron
current density ${\bf j}_{\bf q}=\frac{1}{k_F}\sum_{{\bf k}\sigma}{\bf
  k} c^\dagger_{{\bf k}-{\bf q}/2\sigma}c_{{\bf k}+{\bf q}/2\sigma}$
and a vector mode with the propagator $D_{ij}({\bf
  q},\tau)=\left\langle TA^i_{-{\bf q}}(-i\tau)A_{\bf
  q}^j\right\rangle$ and the Fourier transform $D_{ij}({\bf
  q},\omega_n)=D_\perp({\bf
  q},\omega_n)\left(\delta_{ij}-q_iq_j/q^2\right)$.  In the Nambu
representation we can write
\[
\rho_{\bf q} =\sum_{\bf k}
\alpha^\dagger_{{\bf k}-{\bf q}/2}\tau_3\alpha_{{\bf k}+{\bf q}/2},
\qquad
{\bf j}_{\bf q} =\frac{1}{k_F}\sum_{\bf k} {\bf k}
\alpha^\dagger_{{\bf k}-{\bf q}/2}\tau_0\alpha_{{\bf k}+{\bf q}/2}.
\]
The presence of different $\tau$-matrices in $\rho_{\bf q}$ and ${\bf
  j}_{\bf q}$ is caused by the different behaviour of charge and
current under time reversal.  In the self-consistent Born
approximation, the $2\times 2$ self-energy matrix corresponding to
$H_{\rm int}$ is given by
\begin{eqnarray*}
{\hat\Sigma}_{\bf k}(i\omega_n)&=&
\frac{T}{\cal V}\sum_{{\bf k^\prime}\omega_m}
{\rm g}_{0{\bf k}^\prime-{\bf k}}^2
D_0({\bf k}^\prime-{\bf k},\omega_m-\omega_n)
\tau_3{\cal G}_{\bf k^\prime}(i\omega_m)\tau_3
\\
&+&\frac{T}{\cal V}\sum_{{\bf k^\prime}\omega_m}
\frac{({\bf k}\times {\bf k^\prime})^2}{k_F^2({\bf k^\prime}-{\bf k})^2}
{\rm g}_{\perp{\bf k}^\prime-{\bf k}}^2
D_\perp({\bf k}^\prime-{\bf k},\omega_m-\omega_n)
\tau_0{\cal G}_{\bf k^\prime}(i\omega_m)\tau_0.
\end{eqnarray*}
For the sake of simplicity, in what follows let us replace the factor
$\frac{({\bf k}\times {\bf k^\prime})^2}{k_F^2({\bf k^\prime}-{\bf
    k})^2}$ by its value for forward scattering, i.e. by 1. Performing
furthermore the radial integrations in the usual way by introducing
the electronic density of states in the normal state $N(0)$, the
Eliashberg equations reduce to the standard form
\begin{subequations} 
\label{eq:impure_eliashberg}
\begin{equation}
Z_n=1+\frac{\pi T}{\omega_n}\sum_m \sum_{s=0,\perp}g_s(\omega_n-\omega_m)
\frac{\omega_m}{\sqrt{\omega_m^2+\Delta_m^2}},
\label{eq:impure_eliashberg_a}
\end{equation}
\begin{equation}
Z_n\Delta_n=\pi T \sum_m \sum_{s=0,\perp} \eta_s g_s(\omega_n-\omega_m)
\frac{\Delta_m}{\sqrt{\omega_m^2+\Delta_m^2}},
\label{eq:impure_eliashberg_b}
\end{equation}
\end{subequations}
where $g_s(\omega_n)=\frac{1}{2}N(0)\overline{{\rm
    g}_s^2D_s(\omega_n)}$ are dimensionless coupling functions for the
two modes $s=0,\perp$ and $\overline{{\rm g}_s^2D_s(\omega_n)}$ is a
Fermi-surface average of ${\rm g}_{s{\bf k}^\prime-{\bf k}}^2 D_s({\bf
  k}^\prime-{\bf k},\omega_n)$. 

Note that both modes enter equation~(\ref{eq:impure_eliashberg_a}) for
the wave-function renormalisation in a symmetric way. This is not the
case, however, for equation~(\ref{eq:impure_eliashberg_b}) for the gap
function, since $\eta_0=1$ and $\eta_\perp=-1$, clearly distinguishing
the pairing scalar mode from the pair-breaking vector mode. This
difference is caused by the difference in $\tau$-matrices, i.e. by the
different behaviour of charge and current under time
reversal. Physically, the result $\eta_\perp=-1$ is a consequence of
the repulsive force between antiparallel currents carried by the
electrons forming a Cooper pair.

\begin{figure}
\begin{center}
\begin{minipage}{12cm}
\subfigure[]{
\resizebox*{6.0cm}{!}{\includegraphics{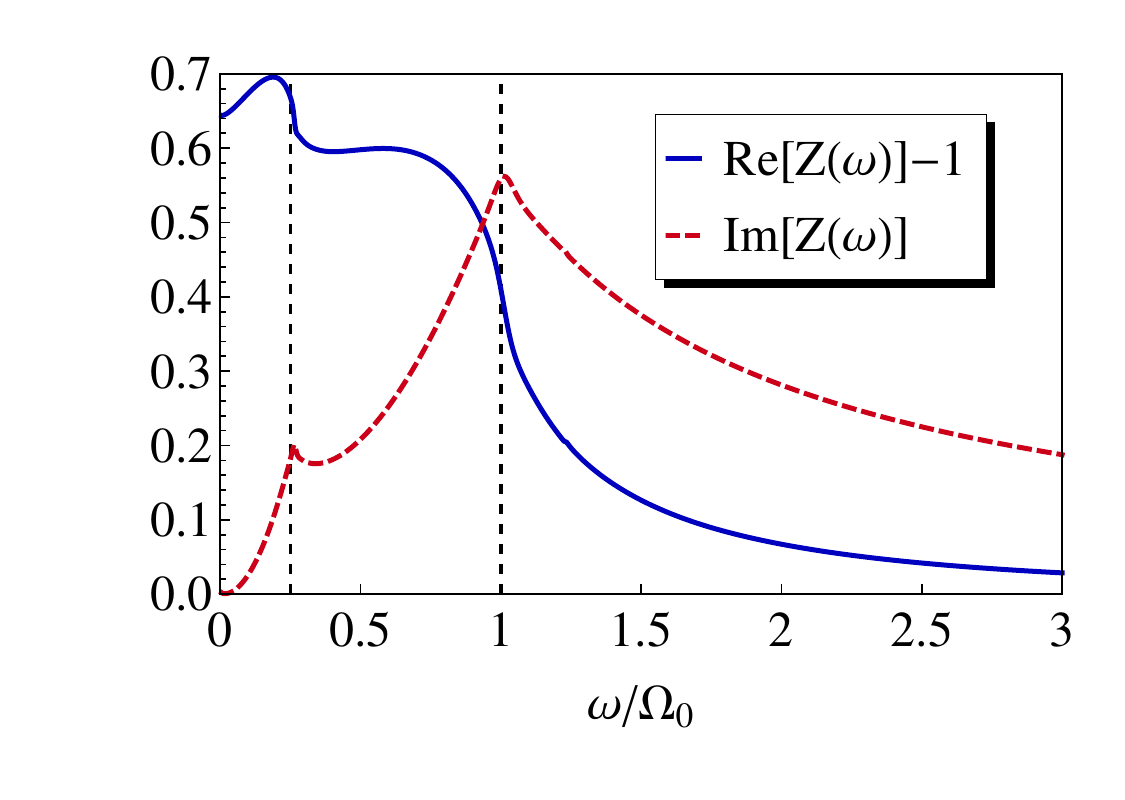}}}
\subfigure[]{
\resizebox*{6.0cm}{!}{\includegraphics{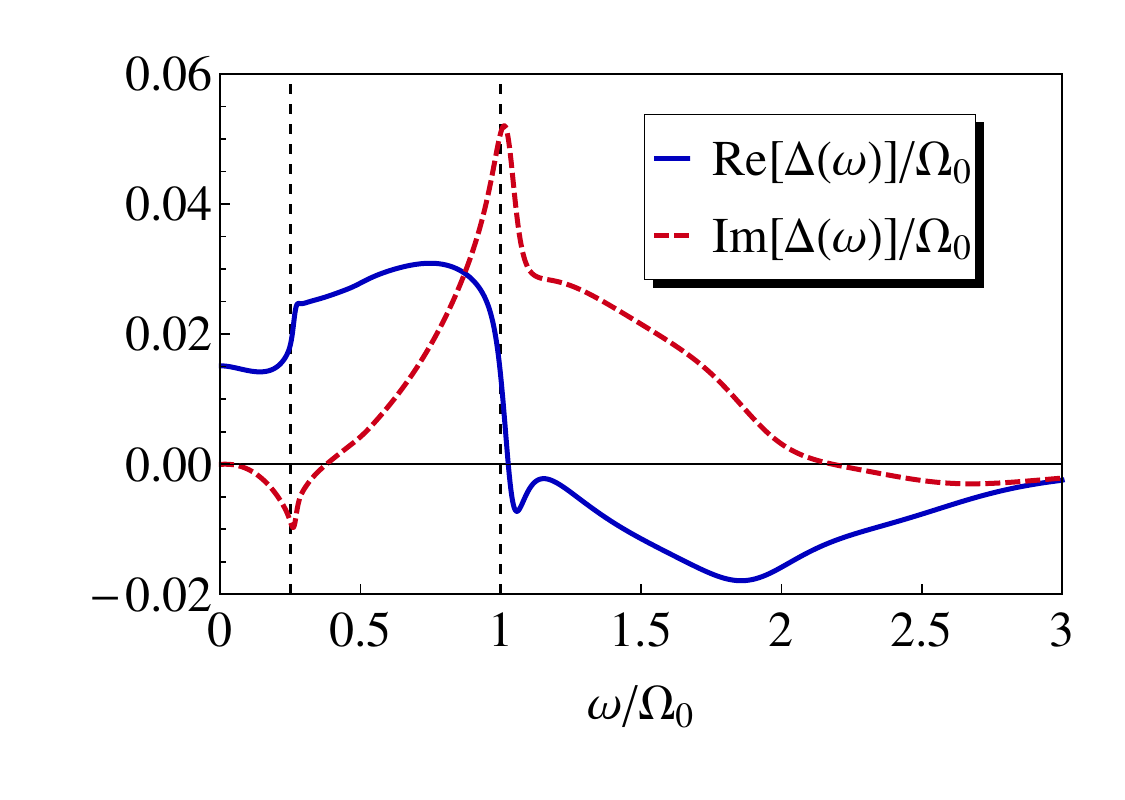}}}
\subfigure[]{
\resizebox*{6.0cm}{!}{\includegraphics{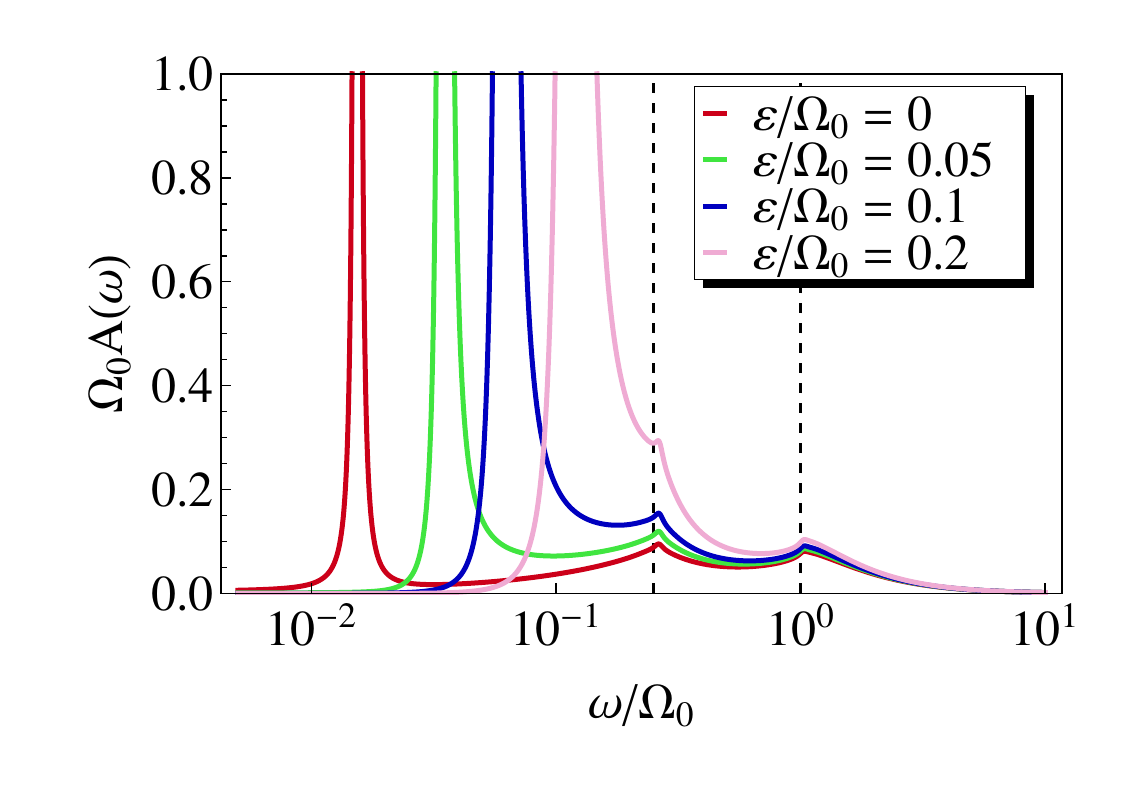}}}
\subfigure[]{
\resizebox*{6.0cm}{!}{\includegraphics{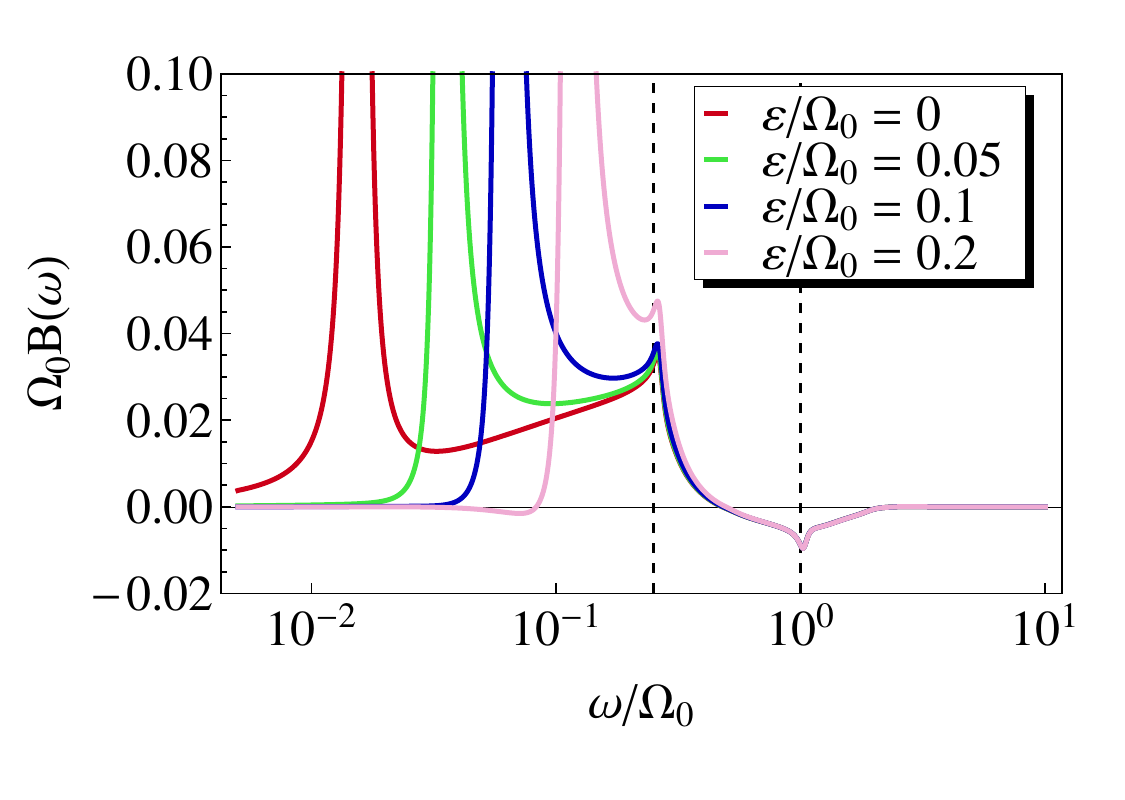}}}
\caption{(colour online) Real and imaginary parts of the wave-function
  renormalisation
  $Z(\omega)=Z^\prime(\omega)+iZ^{\prime\prime}(\omega)$ (a) and of
  the gap function
  $\Delta(\omega)=\Delta^\prime(\omega)+i\Delta^{\prime\prime}(\omega)$
  (b) for the two-mode model~(\ref{eq:model}) with a pairing mode with
  dimensionless coupling $\lambda_0=0.5$ and frequency $\Omega_0$, and
  a pair-breaking mode with $\lambda_\perp=0.15$ and
  $\Omega_\perp=0.25\Omega_0$. The critical temperature of the model
  is $T_c\approx 0.009\Omega_0$. The results were obtained by analytic
  continuation of data obtained on the imaginary axis at temperature
  $T=0.005\Omega_0$. The dotted lines show the energy scales
  $\Omega_0$ and $\Omega_\perp$. Panels (c,d) show the energy
  distribution curves at $\omega>0$ of the normal and anomalous
  spectral functions determined from the data in panels (a,b) at fixed
  values of $\varepsilon_k$ (left to right curves correspond to
  increasing $\varepsilon_k$). Note the logarithmic scale of
  $\omega$.}
\label{fig_2modes}
\end{minipage}
\end{center}
\end{figure}

It is worth pointing out that the Eliashberg
equations~(\ref{eq:impure_eliashberg}) can be generalized to take into
account also finite impurity scattering treated within the
self-consistent Born approximation, if we include an additional
scattering mode $s={\rm imp}$ with an elastic coupling function
$g_{\rm imp}(\omega_n)=\frac{\Gamma_{\rm imp}}{\pi T}\delta_{n0}$ and
$\eta_{\rm imp}=1$. Also a simplified version of pair-breaking elastic
scattering on the magnetic impurities can be described by the same
formalism by including yet another scattering mode $s={\rm mag}$ with
the coupling function $g_{\rm mag}(\omega_n)=\frac{\Gamma_{\rm
    mag}}{\pi T}\delta_{n0}$ and $\eta_{\rm mag}=-1$.

In Figure~\ref{fig_2modes} we plot the wave-function renormalisation
and the gap function, after analytical continuation to the real axis
by means of the Pad\'{e} approximation \cite{Beach00}, for the
two-mode model~(\ref{eq:model}). We have considered coupling to
Debye-type modes with coupling strengths ${\rm g}_{s{\bf q}}\propto
\sqrt{q}$ and linear dispersions $\omega_s(q)=v_sq$, which leads to
the coupling functions $g_s(\omega_n)=\lambda_s
\left[1-\frac{\omega_n^2}{\Omega_{s}^2}
  \ln\left(1+\frac{\Omega_{s}^2}{\omega_n^2}\right)\right]$ with
dimensionless coupling constants $\lambda_s$ and characteristic
frequencies $\Omega_s$.  The crucial observation to be made in
Figure~\ref{fig_2modes} is that the wave-function renormalisation
$Z(\omega)$ does not distinguish between pairing and pair-breaking
modes: for instance, at both characteristic frequencies $\Omega_0$ and
$\Omega_\perp$, the imaginary part $Z^{\prime\prime}(\omega)$ exhibits
local maxima.  Whether we are dealing with a pairing or a
pair-breaking mode is visible only in the gap function
$\Delta(\omega)$. In fact, for $\omega>0$ the imaginary part
$\Delta^{\prime\prime}(\omega)$ exhibits a local minimum close to the
pair-breaking scale $\Omega_\perp$, while close to the pairing scale
$\Omega_0$ it exhibits a maximum. Similarly, when we lower the energy
$\omega$ below $\Omega_0$, the pairing energy $\Delta^\prime(\omega)$
increases, while exactly opposite behaviour is observed in the
vicinity of $\Omega_\perp$.

The same dichotomy is observed also in the language of spectral
functions: the normal spectral function $A_{\bf k}(x)$ does not
distinguish the pairing from the pair-breaking modes, since it
exhibits maxima in the vicinity of both, $\Omega_0$ and
$\Omega_\perp$.  On the contrary, the anomalous spectral function
$B_{\bf k}(x)$ exhibits for $x>0$ local maxima in the vicinity of the
pair-breaking energy $\Omega_\perp$, whereas local minima are observed
close to the pairing energy $\Omega_0$.

Note that in the gauge which we have chosen to work in, $B_{\bf k}(x)$
is real but not positive even if we restrict ourselves to $x>0$.  The
necessity of a sign change of $B_{\bf k}(x)$ follows from the
following exact sum rule which we prove in
Appendix~\ref{sec_appendix_a}:
\begin{equation}
\int_{-\infty}^\infty dx x B_{\bf k}(x)=2\int_0^\infty dx x B_{\bf k}(x)=0.
\label{eq:sum_rule}
\end{equation}
It is worth pointing out that if the Coulomb pseudopotential
$\mu^\ast$ is included in equations~(\ref{eq:impure_eliashberg}) then
the sum rule~(\ref{eq:sum_rule}) does not hold any more, but we have
checked that the function $B_{\bf k}(x)$ still exhibits both, a sign
change and a feature close to the energy of the pairing glue.

\section{Eliashberg theory: determination of the gap function}
\label{sec_dos}

\subsection{Tunneling density of states}
\label{sec_dos1}
It is well known that the tunneling experiments gave the historically
first direct access to the gap function $\Delta(\omega)$ of the
conventional superconductors \cite{Schrieffer64}. It is rarely
stressed, however, that the very possibility to interpret the
tunneling density of states $N_S(\omega)$ solely in terms of
$\Delta(\omega)$ is somewhat surprising. In fact, $N_S(\omega)$ per
spin is given by
\begin{equation}
N_S(\omega)=\frac{1}{\cal V}\sum_{\bf k}A_{\bf k}(\omega)
\approx-\frac{1}{\pi}N(0)\int d\varepsilon_k 
{\rm Im}G_{\bf k}(\omega+i0),
\label{eq:dos_def}
\end{equation}
where in the approximate equality we have assumed that the system is
effectively isotropic, and that the density of states in the normal
state does not vary appreciably on the energy scale of the
superconducting gap. The surprise has to do with the fact that $G_{\bf
  k}(\omega)$ is given by equation~(\ref{eq:green_eliashberg}), that
is by two functions $Z_{\bf k}(\omega)$ and $\Delta_{\bf k}(\omega)$
in the general case. But it should not be possible to determine {\it
  two} functions from a measurement of a {\it single} function
$N_S(\omega)$! The solution to this paradox is that, if $Z_{\bf
  k}(\omega)$ and $\Delta_{\bf k}(\omega)$ do not depend on ${\bf k}$,
the integration over $\varepsilon_k$ can be performed and it leads to
the result (assuming that the branch cut of the square root is set 
along 
the negative real axis)
\begin{equation}
\frac{N_S(\omega)}{N(0)}\approx {\rm Re}
\frac{|\omega|}{\sqrt{\omega^2-\Delta^2(\omega)}},
\label{eq:dos_isotropic}
\end{equation}
which does not depend on $Z(\omega)$ 
any more. Therefore ~(\ref{eq:dos_isotropic}) can be used as a tool 
for measuring
the gap function $\Delta(\omega)$.  It should be pointed out that the
inversion of the formula~(\ref{eq:dos_isotropic}) is not completely
trivial, since
$\Delta(\omega)=\Delta^\prime(\omega)+i\Delta^{\prime\prime}(\omega)$
is a complex function, but thanks to the analytic properties of
$\Delta(\omega)$ it can be performed \cite{Galkin74}.

\begin{figure}
\begin{center}
\begin{minipage}{12cm}
\centerline{\includegraphics[width=8.0cm]{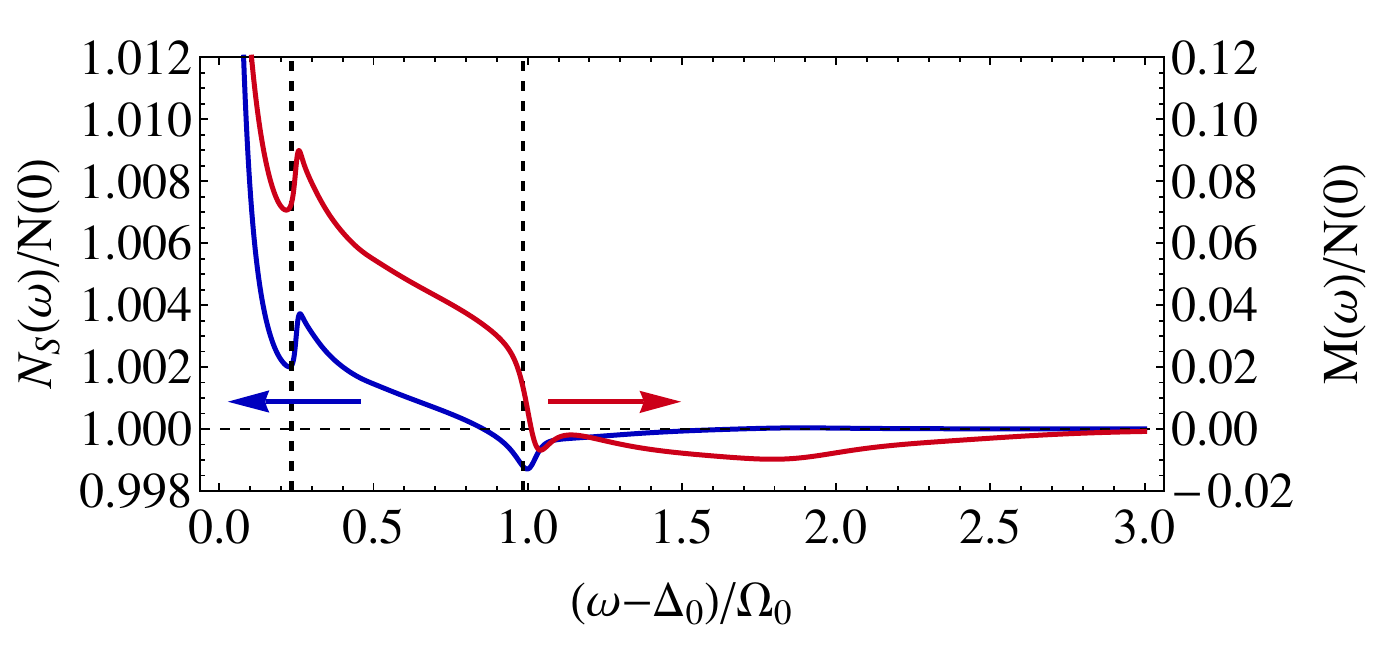}}
\caption{(colour online) Tunneling density of states $N_S(\omega)$ and
  anomalous density of states $M(\omega)$ for the two-mode
  model~(\ref{eq:model}).  The parameters are the same as in
  Fig.~\ref{fig_2modes}.  Note the structure close to the energy
  scales $\Omega_0$ and $\Omega_\perp$ (which are shown by the dotted
  lines).}
\label{fig_dos}
\end{minipage}
\end{center}
\end{figure}

In Figure~\ref{fig_dos} we plot the tunneling density of states
$N_S(\omega)$ for the two-mode model~(\ref{eq:model}).  We have
introduced the spectral gap $\Delta_0$ as a solution to the equation
$\omega=\Delta^\prime(\omega)$ and we find $\Delta_0=0.015\Omega_0$.
Note that $N_S(\omega)$ exhibits local increase at the pair-breaking
scale $\Omega_\perp+\Delta_0$ and a decrease at the pairing scale
$\Omega_0+\Delta_0$.

\subsection{Tomographic density of states}
\label{sec_dos2}
The pairing state within a CuO$_2$ plane is known to be anisotropic
\cite{Tsuei00}, therefore the results for the density of states
derived in Section~\ref{sec_dos1} obviously do not apply to the
cuprates. In order to account for their in-plane anisotropy, let us
introduce the curvilinear coordinates ${\bf k}=(k_\parallel,k_\perp)$,
where $k_\parallel$ varies along the Fermi line and $k_\perp$ varies
along the normal to the Fermi line. Let us further denote the length
of the Fermi line as $\oint dk_\parallel=2\pi k_F$ and instead of
$k_\parallel$ define the angular variable $\theta$ along the Fermi
line by $d\theta=dk_\parallel/k_F$. Let us also replace the coordinate
$k_\perp$ by the non-interacting energy $\varepsilon_k$, making use of
the relation $d\varepsilon_k=v_F(\theta)dk_\perp$, where $v_F(\theta)$
is the Fermi velocity at angle $\theta$.

The functions $Z_{\bf k}(\omega)$ and $\Delta_{\bf k}(\omega)$
entering the definition of the Green's
function~(\ref{eq:green_eliashberg}) depend in general on three
variables $(\varepsilon_k,\theta,\omega)$. If we now assume, in
analogy with conventional superconductors, that their dependence on
$\varepsilon_k$ can be neglected, then the $\varepsilon_k$-integration
in equation~(\ref{eq:dos_def}) can be performed and we arrive at the
expression for the two-dimensional density of states per spin
\begin{equation}
N_S(\omega)=\frac{k_F}{(2\pi)^2}\oint d\theta N_S(\theta,\omega),
\label{eq:dos_anisotropic}
\end{equation}
where we have defined the angle-resolved (tomographic)
density of states
\begin{equation}
N_S(\theta,\omega)=\int_{\theta={\rm const}}dk_\perp A_{\bf k}(\omega)
=\frac{1}{v_F(\theta)}
{\rm Re}\frac{|\omega|}{\sqrt{\omega^2-\Delta^2(\theta,\omega)}}.
\label{eq:dos_tomographic}
\end{equation}

Usually it is assumed that scanning tunneling spectrocopy from the
$c$-axis direction gives us direct access to the tunneling density of
states $N_S(\omega)$ in the cuprates \cite{Berthod13}. However, even
if we assume that
equations~(\ref{eq:dos_anisotropic},\ref{eq:dos_tomographic}) do
apply, their direct inversion with the goal of determining
$\Delta(\theta,\omega)$ is not possible, due to the
$\theta$-dependence of the gap function. The best one can do is to fit
the data for $N_S(\omega)$ to a postulated microscopic model and to
extract the relevant parameters in this way.  A recent example of such
a procedure can be found in \cite{Berthod13}, where the tunneling data
on Bi2223 have been fit to a phenomenological model with
spin-fluctuation induced pairing. It should be pointed out, however,
that no less than 13 parameters were needed in the fits. Clearly, a
less biased technique for determination of $\Delta(\theta,\omega)$ is
needed.

Very recently, it has been realized \cite{Reber12,Reber13} that the
tomographic density of states can be determined from the ARPES data by
taking their integral according to the
definition~(\ref{eq:dos_tomographic}). Very promising results for
$N_S(\theta,\omega)$ have been found for angles $\theta$ lying close
to the nodal direction. Since the authors were interested only in the
simplest questions regarding the interplay between the pair-breaking
rate $\Gamma(\theta)$ and the gap magnitude $\Delta(\theta)$, the
measured tomographic density of states was fitted to the Dynes
formula. However, by applying the powerful inversion technology
\cite{Galkin74} to the data for $N_S(\theta,\omega)$ at fixed angle
$\theta$, it should be possible to obtain the full complex function
$\Delta(\theta,\omega)=\Delta^\prime(\theta,\omega)
+i\Delta^{\prime\prime}(\theta,\omega)$ without making any additional
assumptions about its shape. This might open an unbiased way to the
identification of the pairing glue in the cuprates.

\subsection{Anomalous branch of the momentum distribution curves}
\label{sec_dos3}
Let us assume again that the functions $Z(\theta,\omega)$ and
$\Delta(\theta,\omega)$ depend only on $\theta$ and $\omega$, but not
on $\varepsilon_k$. ARPES measures the spectral function
\begin{equation}
A(\varepsilon_k,\theta,\omega)=-\frac{1}{\pi}{\rm Im}
\left[
\frac{\omega Z(\theta,\omega)+\varepsilon_k}
{\omega^2 Z(\theta,\omega)^2-\phi(\theta,\omega)^2-\varepsilon_k^2}
\right],
\label{eq:momentum_distrib}
\end{equation}
where $\phi(\theta,\omega)=Z(\theta,\omega)\Delta(\theta,\omega)$ is
the anomalous self-energy.  It is well established \cite{Campuzano04}
that, in the normal state and in absence of a gap, the spectral
function $A(\varepsilon_k,\theta,\omega)$ of the cuprates exhibits a
simple Lorentzian peak when plotted as a function of $\varepsilon_k$
at fixed values of $\omega$ and $\theta$.  Such plots are called
momentum distribution functions and their simple shape can be taken as
an a posteriori evidence for independence of $Z(\theta,\omega)$ on
$\varepsilon_k$ \cite{Campuzano04}. Moreover, if the bare dispersion
$\varepsilon_k$ is known, then the function
$Z(\theta,\omega)=Z^\prime(\theta,\omega)+iZ^{\prime\prime}(\theta,\omega)$
can be easily determined by fitting the momentum distribution
functions to a Lorentzian.

Let us generalize the predictions of
equation~(\ref{eq:momentum_distrib}) for the momentum distribution
function to the superconducting state.  To illustrate the point, let
us consider the simplest case with real $Z(\theta,\omega)$ and
$\Delta(\theta,\omega)$. In that case we find
\[
A(\varepsilon_k,\theta,\omega)=
\frac{|\omega|+{\rm sgn}(\omega)\Omega}
{2\Omega}\delta(\varepsilon_k-Z\Omega)
+\frac{|\omega|-{\rm sgn}(\omega)\Omega}
{2\Omega}\delta(\varepsilon_k+Z\Omega),
\]
where $\Omega(\theta,\omega)=\sqrt{\omega^2-\Delta^2(\theta,\omega)}$.
This result shows, as is well known, that excitations with a given
energy $\omega$ are realized at two momenta: one inside and another
one outside the Fermi sea, corresponding to $\varepsilon_k=\pm
Z(\theta,\omega)\Omega(\theta,\omega)$. If only one of the branches is
observable, then the measurement of the quasiparticle dispersion can
not be inverted to obtain the function $\Delta(\theta,\omega)$, since
the wave-function renormalisation $Z(\theta,\omega)$ is unknown.
However, if both branches are observable, then the additional
information on the relative weight of the two branches allows us to
determine $\Omega(\theta,\omega)$, and, as a consequence, both
$\Delta(\theta,\omega)$ and $Z(\theta,\omega)$ can be determined.

\begin{figure}
\begin{center}
\begin{minipage}{12cm}
\centerline{\includegraphics[width=8.0cm]{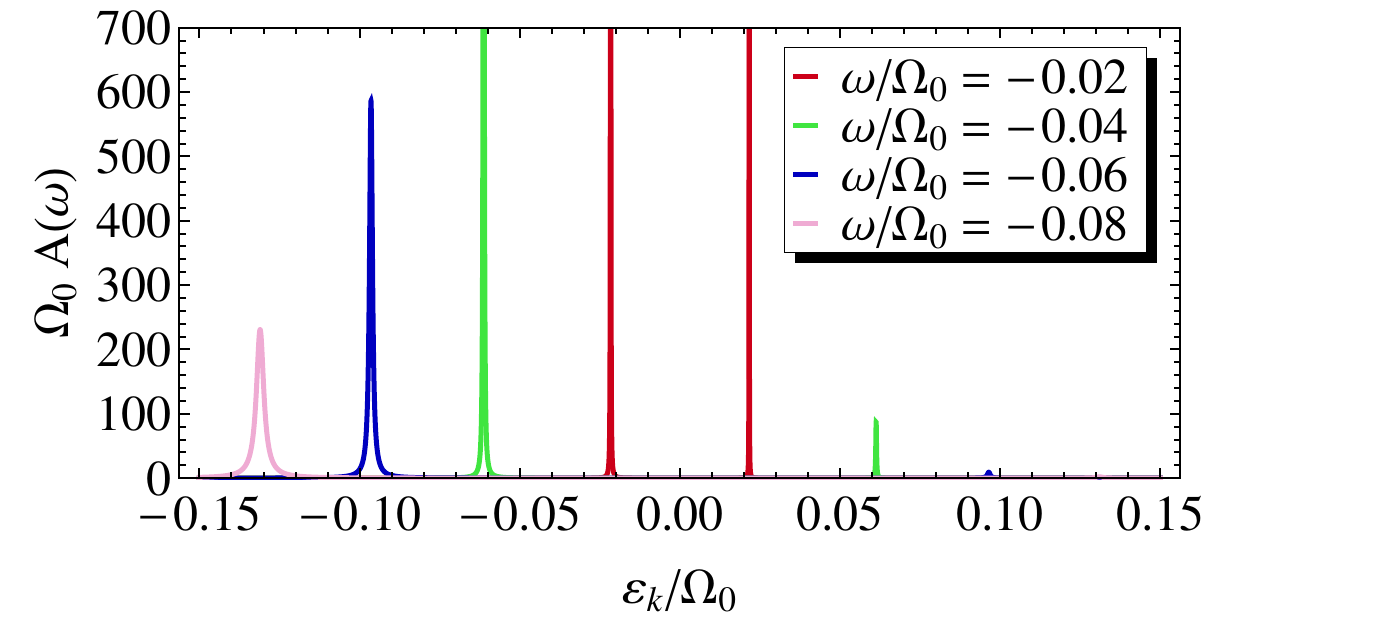}}
\caption{(colour online) Momentum distribution curves for the two-mode
  model~(\ref{eq:model}). The parameters are the same as in
  Fig.~\ref{fig_2modes}. The curves consist of two peaks placed
  symmetrically around the Fermi surface $\varepsilon_k=0$. Their
  distance from $\varepsilon_k=0$ increases with increasing
  $|\omega|$.}
\label{fig_mdc}
\end{minipage}
\end{center}
\end{figure}

In a sense, the observation of the two branches gives us two
independent measurements which enable the determination of the two
independent functions $\Delta(\theta,\omega)$ and $Z(\theta,\omega)$.
This idea has been used very recently in an interesting paper
\cite{Zhang12}, where the momentum distribution curves for a set of
angles $\theta$ have been studied on a fine grid of energies
$\omega<0$. At fixed $\omega$ and $\theta$, the data was fitted to
equation~(\ref{eq:momentum_distrib}), making use of 4 parameters
describing the real and imaginary parts of the wave-function
renormalisation $Z=Z^\prime+iZ^{\prime\prime}$ as well as of the
anomalous self-energy $\phi=\phi^\prime+i\phi^{\prime\prime}$. For
technical reasons, two additional fitting parameters had to be used:
one for the overall intensity, and another one for background
subtraction. The bare dispersion $\varepsilon_k$ was taken from a
tight-binding model.

The momentum distribution curves for the two-mode
model~(\ref{eq:model}) for several values of $\omega$ are shown in
Figure~\ref{fig_mdc}. Note that the anomalous branch has substantial
weight only for those energies $\omega$, for which
$|\omega|\sim\Delta_0$. For larger values of $|\omega|$, the
results for the anomalous self-energy necessarily become quite noisy,
as in fact found also in \cite{Zhang12}. Therefore this technique is
potentially useful in cases where the characteristic energy of the
pairing glue is comparable to the energy gap.

\section{Direct measurement of the anomalous spectral function}
\label{sec_anomalous}
All of the techniques for determination of the gap function
$\Delta_{\bf k}(\omega)$ described in Section~\ref{sec_dos} were based
on the measurement of the diagonal spectral function $A_{\bf
  k}(x)$. The crucial feature which had to be assumed was that the
functions $\Delta(\varepsilon_k,\theta,\omega)$ and
$Z(\varepsilon_k,\theta,\omega)$ do not depend on
$\varepsilon_k$. While this assumption is well established in the
context of low-temperature superconductors, it is far from obvious
that it applies to the cuprates as well. On one hand, the simplicity
of the momentum distribution curves as opposed to the complicated
energy distribution curves in ARPES can be taken as indirect evidence
for such weak $\varepsilon_k$-dependence \cite{Campuzano04}.  But on
the other hand, since the cuprates are doped Mott insulators, one
should expect large particle-hole asymmetry, and therefore also strong
$\varepsilon_k$-dependence, as in fact observed in tunneling
experiments \cite{Pushp09}. Also the more standard concept of
spin-fluctuation mediated superconductivity naturally leads to a
strong $\varepsilon_k$-dependence, because only at the hot spots the
scattering is equally strong for electron states inside and outside
the Fermi sea. Away from the hot spots, scattering either inside or
outside the Fermi sea should dominate, depending on the angle
$\theta$.

An obvious recipe to determine the {\it two functions} $\Delta_{\bf
  k}(\omega)$ and $Z_{\bf k}(\omega)$ is to measure the {\it two
  spectral functions} characterizing a superconductor, namely $A_{\bf
  k}(x)$ and $B_{\bf k}(x)$. We are thus led to the search for a
suitable spectroscopic technique for the anomalous spectral function
$B_{\bf k}(x)$. A natural candidate is the Josephson effect, where the
Cooper pairs enter or leave a given superconductor. However, the
Josephson current is determined by an integral of $B_{\bf k}(x)$,
providing just one number and not a function of a continuous
parameter, needed for a spectroscopic measurement. We are therefore
led to the study of the frequency spectrum of temporal fluctuations of
the current $I(t)$ traversing the Josephson junction at zero bias
$S(\omega)=\int_{-\infty}^\infty dt \langle I(t) I(0)\rangle
e^{i\omega t}$ or, making use of the fluctuation-dissipation theorem
$S(\omega)=-2\left[n(\omega)+1\right]\chi^{\prime\prime}(\omega)$, of
the retarded current-current correlation function
\begin{equation}
\chi(\omega)=-i\int_{-\infty}^\infty dt e^{i(\omega+i0) t}
\langle [I(t),I(0)]\rangle \theta(t).
\label{eq:current_current}
\end{equation}

In what follows we calculate $\chi^{\prime\prime}(\omega)$ for a
Josephson junction between two superconductors, which we call left and
right.  Let us denote the single-particle states of the left and right
superconductors by ${\bf k}$ and ${\bf q}$, respectively. If we
describe the junction by the tunneling Hamiltonian $H_{\rm
  tun}=\sum_{{\bf k}{\bf q}} \left[t_{{\bf k}{\bf q}}(c^\dagger_{{\bf
      k}\uparrow}c_{{\bf q}\uparrow} +c^\dagger_{-{\bf
      q}\downarrow}c_{-{\bf k}\downarrow})+{\rm h.c.}  \right]$, the
current operator is given by $I=-ie\sum_{{\bf k}{\bf q}} \left[t_{{\bf
      k}{\bf q}}\alpha^\dagger_{{\bf k}}\alpha_{{\bf q}}-{\rm
    h.c.}\right]$.  A simple evaluation of~(\ref{eq:current_current})
to second order in the tunneling matrix element gives
\begin{equation}
\chi^{\prime\prime}(\omega)=2\pi e^2\sum_{{\bf k}{\bf q}}|t_{{\bf k}{\bf q}}|^2
\int_{-\infty}^\infty dx \left[f(x+\omega)-f(x)\right]
\left[A_{{\bf k}{\bf q}}(x,\omega)+B_{{\bf k}{\bf q}}(x,\omega)\right],
\label{eq:chi_ns}
\end{equation}
where $A_{{\bf k}{\bf q}}(x,\omega)= A_{\bf k}(x)A_{\bf q}(x+\omega)
+A_{\bf k}(-x)A_{\bf q}(-x-\omega)$ generates the normal contribution
$\chi_N^{\prime\prime}(\omega)$ and $B_{{\bf k}{\bf q}}(x,\omega)=
B_{\bf k}(x)B_{\bf q}^\ast(x+\omega) +B_{\bf k}^\ast(x)B_{\bf
  q}(x+\omega)$ generates the superconducting contribution
$\chi_S^{\prime\prime}(\omega)$.

If we drive one of the superconductors forming the junction into its
normal state, e.g. by applying magnetic field or by heating above
$T_c$, we can determine the normal contribution
$\chi_N^{\prime\prime}(\omega)$ and therefore isolate the
superconducting contribution
$\chi_S^{\prime\prime}(\omega)=\chi^{\prime\prime}(\omega)
-\chi_N^{\prime\prime}(\omega)$. Alternatively, we can study the
current fluctuations around a finite supercurrent stabilized by a
phase difference $\phi$ between the electrodes. In that case we find
$\chi_S^{\prime\prime}(\omega)=\chi_{S0}^{\prime\prime}(\omega)\cos\phi$,
where
\[
\chi_{S0}^{\prime\prime}(\omega)=4\pi e^2\sum_{{\bf k}{\bf q}}|t_{{\bf k}{\bf q}}|^2
\int_{-\infty}^\infty dx \left[f(x+\omega)-f(x)\right]
B_{\bf k}(x)B_{\bf q}(x+\omega),
\]
and the anomalous spectral functions are calculated in a gauge where
both superconductors have a real order parameter.  Measuring the
component of $S(\omega)$ proportional to $\cos\phi$, we can therefore
directly measure $\chi_{S0}^{\prime\prime}(\omega)$.

In order to proceed, let us further assume that the tunneling is
featureless, $|t_{{\bf k}{\bf q}}|^2=|t|^2{\cal S}/{\cal V}^2$, where
${\cal S}$ is the junction area and ${\cal V}$ is the normalisation
volume. This is a standard assumption for tunnel junctions with an
oxide barrier \cite{Schrieffer64}, but it should apply also to
scanning tunneling spectrocopy with a superconducting tip
\cite{Berthod13}.  If we furthermore introduce the anomalous versions
of the tunneling density of states for the left and right
superconductors, $M_L(\omega)=\frac{1}{\cal V}\sum_{\bf k}B_{\bf
  k}(\omega)$ and $M_R(\omega)=\frac{1}{\cal V}\sum_{\bf q}B_{\bf
  q}(\omega)$, then the function $\chi_{S0}^{\prime\prime}(\omega)$ can
be written as
\begin{equation}
\chi_{S0}^{\prime\prime}(\omega)=4\pi e^2|t|^2 {\cal S}
\int_{-\infty}^\infty dx \left[f(x+\omega)-f(x)\right]
M_L(x)M_R(x+\omega).
\label{eq:chi_s0}
\end{equation}
Since the critical current of the junction is given by
$I_c=\frac{1}{\pi e}\int_{-\infty}^\infty
\frac{d\omega}{\omega}\chi_{S0}^{\prime\prime}(\omega)$, the typical
fluctuating current is $\langle I^2\rangle\sim \frac{e}{\tau}I_c$,
where $\tau$ is a typical fluctuation time defined by
$\chi_{S0}^{\prime\prime}(\omega)$. Note that $\langle I^2\rangle$
scales with the junction area ${\cal S}$, as should have been
expected.

We shall demonstrate the usefulness of equation~(\ref{eq:chi_s0}) by
applying it to the conventional low-temperature superconductors, in
which case the functions $Z$ and $\Delta$ do not depend on
momentum. From equation~(\ref{eq:green_eliashberg}) it then follows
that the anomalous density of states is
\begin{equation}
\frac{M(\omega)}{N(0)}=-\frac{1}{\pi}
\int d\varepsilon_k {\rm Im} F_{\bf k}(\omega+i0)
={\rm Re}\frac{{\rm sgn}(\omega)\Delta(\omega)}
{\sqrt{\omega^2-\Delta^2(\omega)}}.
\label{eq:dos_anomalous}
\end{equation}
Note the similarity of the result for $M(\omega)$ to
equation~(\ref{eq:dos_isotropic}), in particular that neither
$M(\omega)$, nor $N_S(\omega)$ depend on the wave function
renormalisation $Z(\omega)$. Therefore both functions can be used as a
direct measure of $\Delta(\omega)$.  An explicit example of the
anomalous density of states for the two-mode model~(\ref{eq:model}) is
shown in Figure~\ref{fig_dos}.  Note that $M(\omega)$ carries the same
spectroscopic information as $N_S(\omega)$, but its spectroscopic
features are (i) more pronounced and (ii) not masked by a large
constant background $N(0)$.

\begin{figure}
\begin{center}
\begin{minipage}{12cm}
\centerline{\includegraphics[width=8.0cm]{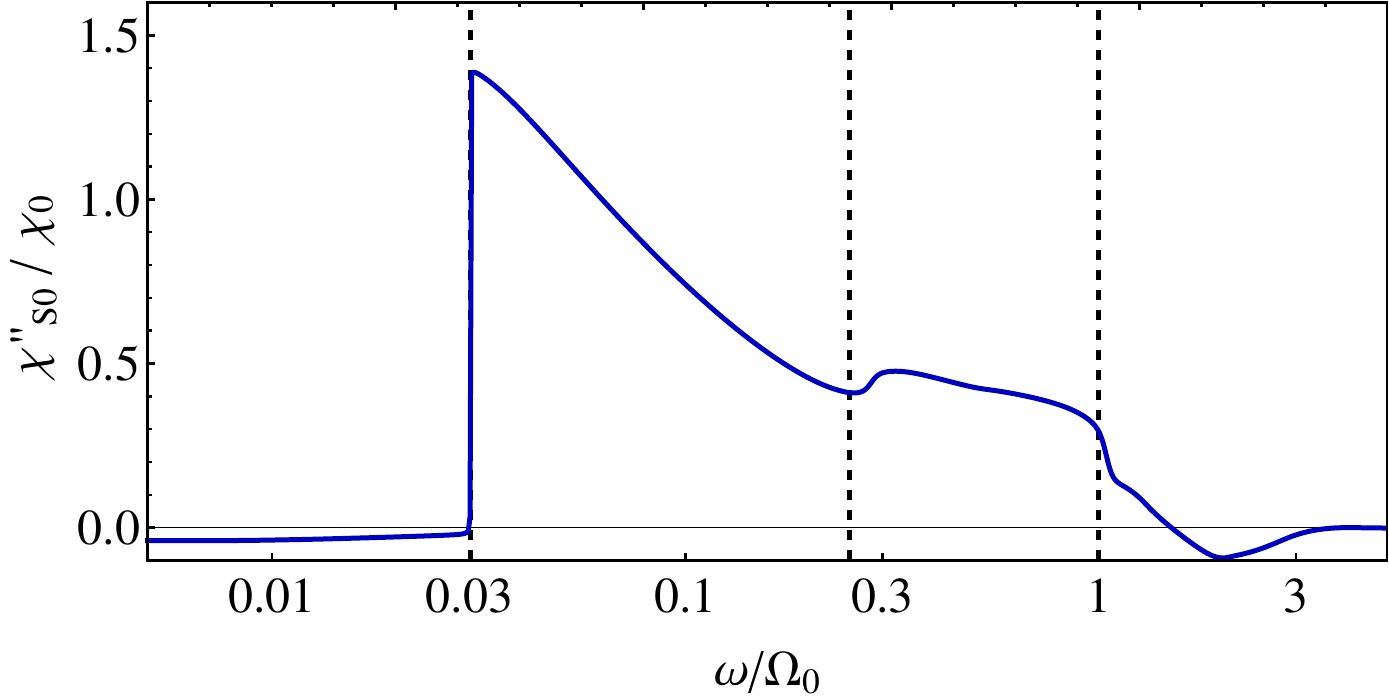}}
\caption{(colour online) The current-current correlation function
  $\chi_{S0}^{\prime\prime}(\omega)$ in units of $\chi_0=4\pi e^2|t|^2
  {\cal S}N(0)^2\Delta_0$.  We assume that $M_L(\omega)=M_R(\omega)$
  and for the anomalous density of states we take the result for the
  two-mode model~(\ref{eq:model}) shown in Fig.~\ref{fig_dos}. Dotted
  lines show the energy scales $2\Delta_0$, $\Omega_\perp$ and
  $\Omega_0$. Note the logarithmic scale of $\omega$.}
\label{fig_chi}
\end{minipage}
\end{center}
\end{figure}

In Figure~\ref{fig_chi} we plot the function
$\chi_{S0}^{\prime\prime}(\omega)$ for a junction formed by two
identical superconductors described by the two-mode
model~(\ref{eq:model}).  As expected, the pairing and pair-breaking
scales are discriminated by $\chi_{S0}^{\prime\prime}(\omega)$ in the
same way as by $M(\omega)$.

Let us conclude by suggesting a two-step procedure for quantitative
determination of the pairing glue from the measured function
$\chi_{S0}^{\prime\prime}(\omega)$.

In the first step, equation~(\ref{eq:chi_s0}) needs to be inverted and
the function $M(\omega)$ of the studied superconductor (say, the right
one) has to be found. This task can be reduced to matrix inversion in
the case when the function $M_L(\omega)$ is known, i.e. when one of
the superconductors forming the Josephson junction is well
understood. Alternatively, in Appendix~\ref{sec_appendix_b} we suggest
a low-temperature inversion procedure for the case when both sides of
the junction are made of the studied superconductor.

In the second step, we should invert~(\ref{eq:dos_anomalous}) to
obtain the complex function $\Delta(\omega)$. It should be possible to
perform this task by modifying the approach of \cite{Galkin74}.

\section{Conclusions}
\label{sec_conclusions}
To summarize, we have presented a short review of the techniques used
to extract information on the pairing glue in superconductors, having
in mind their applicability to the cuprates. All currently available
experimental techniques exploit data on the normal spectral function
$A_{\bf k}(x)$.  The most promising approaches of this type have been
described in Sections~\ref{sec_dos2} and \ref{sec_dos3}. Since all
$A_{\bf k}(x)$-based techniques need to postulate that the normal and
anomalous self-energies do not depend on $\varepsilon_k$ and since
this assumption may not hold in the cuprates, in
Section~\ref{sec_anomalous} we have suggested a novel technique for a
direct measurement of the anomalous spectral function $B_{\bf k}(x)$,
which makes use of the correlation function $\chi(\omega)$ of currents
in a Josephson junction involving the studied superconductor.  This
enables in principle a much more direct determination of the pairing
glue in a superconductor, when compared to the conventional $A_{\bf
  k}(x)$-based techniques.

As a proof of concept, we have analyzed $\chi(\omega)$ in junctions
between conventional isotropic superconductors in the simplest case of
featureless tunneling. Obviously, in order to be applicable to the
cuprates, some degree of directionality of tunneling will need to be
allowed for, since in a $d$-wave superconductor the anomalous density
of states $M(\omega)$ vanishes. Moreover, especially in junctions
involving underdoped cuprates, the effect of phase fluctuations on
$\chi(\omega)$ may have to be included. The most pressing question,
however, is how to measure $\chi(\omega)$ at the (presumably) high
typical frequencies of the pairing glue, when the complex conductivity
of the junction is likely to be dominated by its capacitance.

Quite generally, irreducible correlation functions of other
observables, such as charge-charge, spin-spin, etc., involve normal
and superconducting contributions
resembling~(\ref{eq:chi_ns}). Unfortunately, their analysis is much
more complicated than in the present case, because of the
non-vanishing vertex corrections.  Moreover, even in simplest cases
the physical correlation functions are given as a geometric series of
irreducible correlation functions. Nevertheless, it might be possible
to extract $B_{\bf k}(x)$ also from appropriate combinations of
correlation functions which are more readily measurable than
$\chi(\omega)$. Such approaches definitely deserve further study.

\section*{Acknowledgement}
This work was supported by the Slovak Research and Development Agency
under Grant No.~APVV-0558-10.

\appendices
\section{Proof of the sum rule~(\ref{eq:sum_rule})}
\label{sec_appendix_a}
Because of the symmetry $\phi_{\bf
  k}^{\prime\prime}(\omega)=-\phi_{\bf k}^{\prime\prime}(-\omega)$ of
the anomalous self-energy $\phi_{\bf k}(\omega)=Z_{\bf
  k}(\omega)\Delta_{\bf k}(\omega)$, the Kramers-Kronig relations
imply in the limit $\omega\rightarrow\infty$ the scaling
\[
\phi_{\bf k}^{\prime}(\omega)-\phi_{\bf k}^{\prime}(\infty)=
\frac{1}{\pi}\int_{-\infty}^\infty \frac{dz \phi_{\bf k}^{\prime\prime}(z)}{z-\omega}
\propto\frac{1}{\omega^2}.
\]
Since in any dynamic theory without instantaneous interactions
$\phi_{\bf k}^{\prime}(\infty)=0$, in the limit $\omega\rightarrow\infty$ we
therefore have $\phi_{\bf k}^{\prime}(\omega)\propto\frac{1}{\omega^2}$ and
from equation~(\ref{eq:green_eliashberg}) it follows that
$F_{\bf k}^\prime(\omega)\propto \frac{1}{\omega^4}$. On the other hand, the
spectral representation~(\ref{eq:spectral_representation}) can be
written at $\omega\rightarrow\infty$ as a power series
\[
F_{k}^\prime(\omega)=\sum_{n=0}^\infty\frac{1}{\omega^{n+1}}
\int_{-\infty}^\infty dx x^n B_{\bf k}(x).
\]
But because $F_{\bf k}^\prime(\omega)\propto \frac{1}{\omega^4}$, the
first three terms $n=0,1,2$ have to vanish. Since $B_{\bf k}(x)$ is
odd, the results for $n=0$ and 2 are trivial. On the other hand, the
result for $n=1$ is the sum rule~(\ref{eq:sum_rule}).

\section{Inversion of~(\ref{eq:chi_s0}) in the case $M_L(x)=M_R(x)$}
\label{sec_appendix_b}
The functions $\chi''_{S0}(\omega)$ and $M_L(\omega)=M_R(\omega)\equiv
M(\omega)$ are antisymmetric, so they can be reconstructed from their
values at $\omega>0$. We therefore define
$\widetilde{\chi}(\omega)=\chi''_{S0}(\omega)\theta(\omega)$ and
$\widetilde{M}(\omega)=M(\omega)\theta(\omega)$, where $\theta(\omega)$
is the step function.  The crucial observation is that at zero
temperature
$\left[f(x+\omega)-f(x)\right]\theta(\omega)=-\theta(-x)\theta(x+\omega)$,
and therefore
\[
\widetilde{\chi}(\omega)=
4e^2\left|t\right|^2\mathcal{S}\int_{-\infty}^{\infty} dx 
\widetilde{M}(-x)\widetilde{M}(x+\omega).
 \]
Since $\widetilde{\chi}(\omega)$ is a convolution, Fourier
transformation to the time domain leads to $\widetilde{\chi}(t)\propto
\widetilde{M}^2(t)$.  The inversion of~(\ref{eq:chi_s0}) is now
straightforward. One first finds the Fourier transform
$\widetilde{\chi}(t)$, then takes the square root to obtain
$\widetilde{M}(t)$, Fourier transforms back to the frequency domain,
and finally reconstructs $M(\omega)$ from $\widetilde{M}(\omega)$. The
second step is subtle because the square root has two branches. The
procedure is feasible only if $\widetilde{M}(t)$ is a {\it continuous}
function that {\it never crosses zero}. However, the second condition
fails only in very special cases, because $\widetilde{M}(t)$ is a
complex-valued function.


\label{lastpage}

\end{document}